\documentclass[journal,final]{IEEEtran}
\ifCLASSINFOpdf
  \usepackage[pdftex]{graphicx}
  \graphicspath{{figures/}}
	\usepackage{xcolor}
	\usepackage{transparent}
\else
\fi
\usepackage{arydshln} 
\usepackage{nidanfloat} 
\usepackage{amsmath,amssymb,amsthm, amsfonts}
\makeatletter
\let\oldtheequation\theequation
\renewcommand\tagform@[1]{\maketag@@@{\ignorespaces#1\unskip\@@italiccorr}}
\renewcommand\theequation{(\oldtheequation)}
\makeatother

\DeclareMathOperator{\Max}{Max}  
\DeclareMathOperator{\Min}{Min}  
\DeclareMathOperator{\MSE}{MSE}  
\DeclareMathOperator{\SSIM}{SSIM}  
\newcommand{\density}{\ensuremath{\widetilde{D}}} 
\newcommand{\p}{\density} 
\newcommand{\densityPerSlice}{\ensuremath{D}} 
\newcommand{\numSamples}{\ensuremath{N}} 
\newcommand{\N}{\numSamples} 
\newcommand{\disparity}{\ensuremath{d}} 
\newcommand{\normDisp}{\ensuremath{\hat{d}}} 
\newcommand{\nd}{\normDisp} 
\newcommand{\occluder}{\ensuremath{o}} 
 
\newcommand{\aperture}{\ensuremath{a}} 
\newcommand{\volumeHeight}{\ensuremath{l}} 
\newcommand{\visibility}{\ensuremath{V}} 
\newcommand{\normVisibility}{\ensuremath{\widehat{V}}} 
\newcommand{\nV}{\normVisibility} 
\newcommand{\baseline}{\ensuremath{b}} 
\newcommand{\radius}{\aperture} 
\newcommand{\Var}{\operatorname{Var}}
\newcommand{\Cov}{\operatorname{Cov}}
\newcommand{\dprob}{\ensuremath{q}} 
\newcommand{\q}{\dprob} 
\newcommand{\E}{\operatorname{E}}

\DeclareMathSymbol{\ast}{\mathbin}{symbols}{"03}

\ifCLASSOPTIONcompsoc
  \usepackage[caption=false,font=normalsize,labelfont=sf,textfont=sf]{subfig}
\else
  \usepackage[caption=false,font=footnotesize]{subfig}
\fi
\usepackage{float}
\usepackage{url} 
\usepackage{hyperref} 
\usepackage{siunitx} 

\begin{document}
\title{A Statistical View on Synthetic Aperture Imaging for Occlusion Removal}

\author{Indrajit~Kurmi,
        David~C.~Schedl,
        and~Oliver~Bimber *
				\thanks{Manuscript received March 26; revised May 3; accepted June 10, 2019.}
				\thanks{* Johannes Kepler University Linz, e-mail: first.lastname@jku.at.}
}

\maketitle

\begin{abstract}
Synthetic apertures find applications in many fields, such as radar, radio telescopes, microscopy, sonar, ultrasound, LiDAR, and optical imaging. 
They approximate the signal of a single hypothetical wide aperture sensor with either an array of static small aperture sensors or a single moving small aperture sensor. 
Common sense in synthetic aperture sampling is that a dense sampling pattern within a wide aperture is required to reconstruct a clear signal. 
In this article we show that there exists practical limits to both, synthetic aperture size and number of samples for the application of occlusion removal. 
This leads to an understanding on how to design synthetic aperture sampling patterns and sensors in a most optimal and practically efficient way. 
We apply our findings to airborne optical sectioning which uses camera drones and synthetic aperture imaging to computationally remove occluding vegetation or trees for inspecting ground surfaces.
\end{abstract}

\IEEEpeerreviewmaketitle

\section{Introduction}\label{sec:intro}
Synthetic apertures (SA) approximate the signal of a single hypothetical wide aperture sensor with either an array of static small aperture sensors or a single moving small aperture sensor whose individual signals are computationally combined to increase resolution, depth-of-field, frame rate, contrast, and signal-to-noise ratio. 

This principle has been used in many fields, such as for synthetic aperture radar (SAR) \cite{Moreira2013,Li2015,Rosen2000}, synthetic aperture radio telescopes (SART) \cite{Levanda2010,Dravins2015}, interferometric synthetic aperture microscopy (ISAM) \cite{Ralston2007}, synthetic aperture sonar (SAS) \cite{Hayes2009,Hansen2011}, synthetic aperture ultrasound (SAU) \cite{Jensen2006Review,Zhang2016}, and synthetic aperture LiDAR (SAL) / synthetic aperture imaging laser (SAIL) \cite{Barber2014,Turbide2017}. 
 
In the visible range, synthetic aperture imaging (SAI) \cite{Vaish04,Vaish06,Zhang2018,YangTao2016,Joshi2007,ZhaoPei2019,YangTao2014,ZhaoPei2013} has been used together with large camera arrays that capture structured light fields (regularly sampled multiscopic scene representations) and enable the computation of virtual views with maximal synthetic apertures that correspond to the physical size of the applied camera array. The wide aperture signal results in a shallow depth of field and consequently in a strong blur of out-of-focus occluders, while images of points in focus remain clearly visible. Shifting focus computationally allows optical slicing through dense occluder structures (such as bushes, leaves, tree branches, and coniferous trees), and discovery and inspection of concealed artefacts or objects behind the occluders. 

\begin{figure*}[hb]%
	\centering%
\subfloat[][]{ 
	{\includegraphics[width=0.305\linewidth]{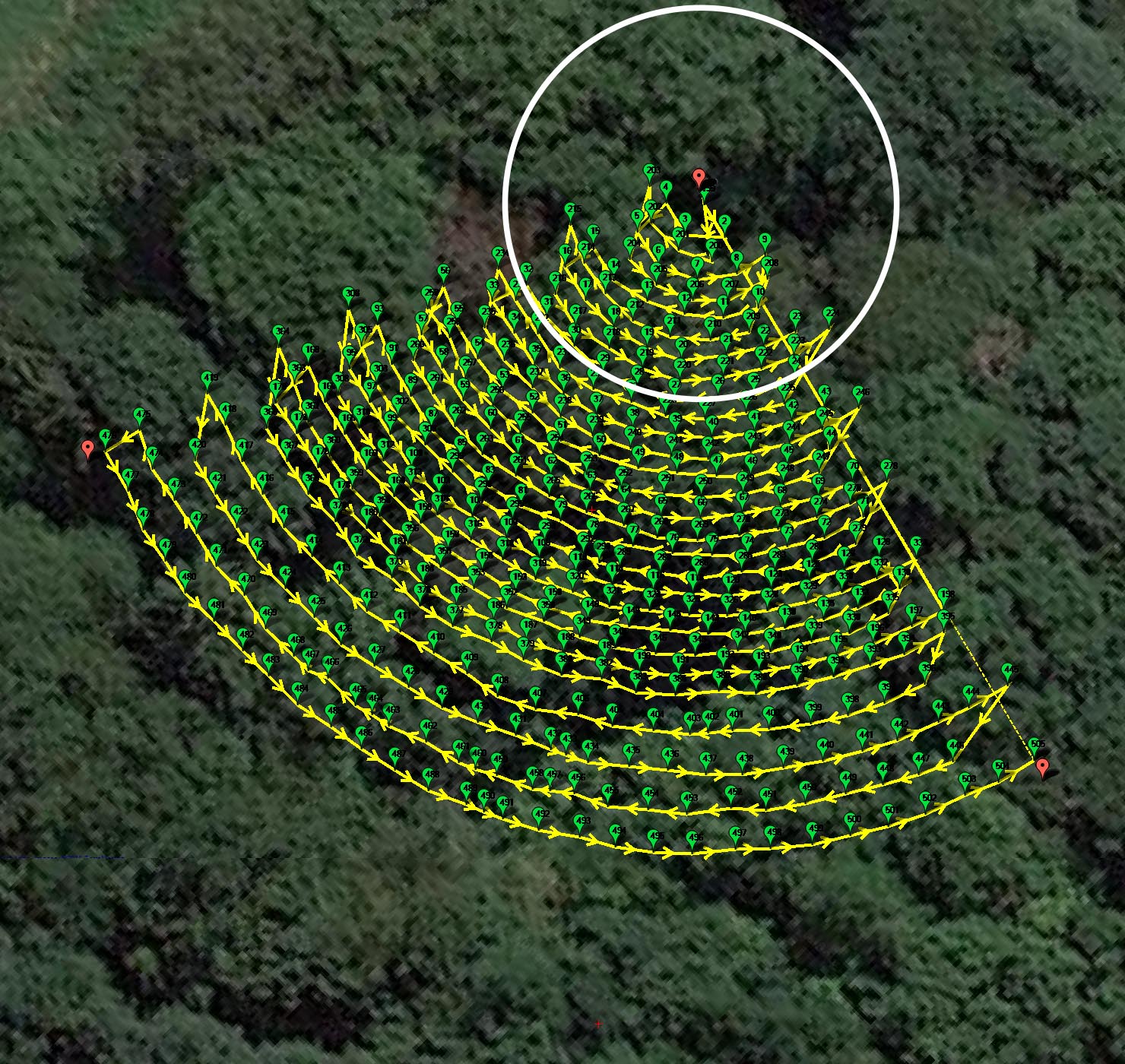}}
}%
\subfloat[][]{ 
	{\includegraphics[width=0.305\linewidth]{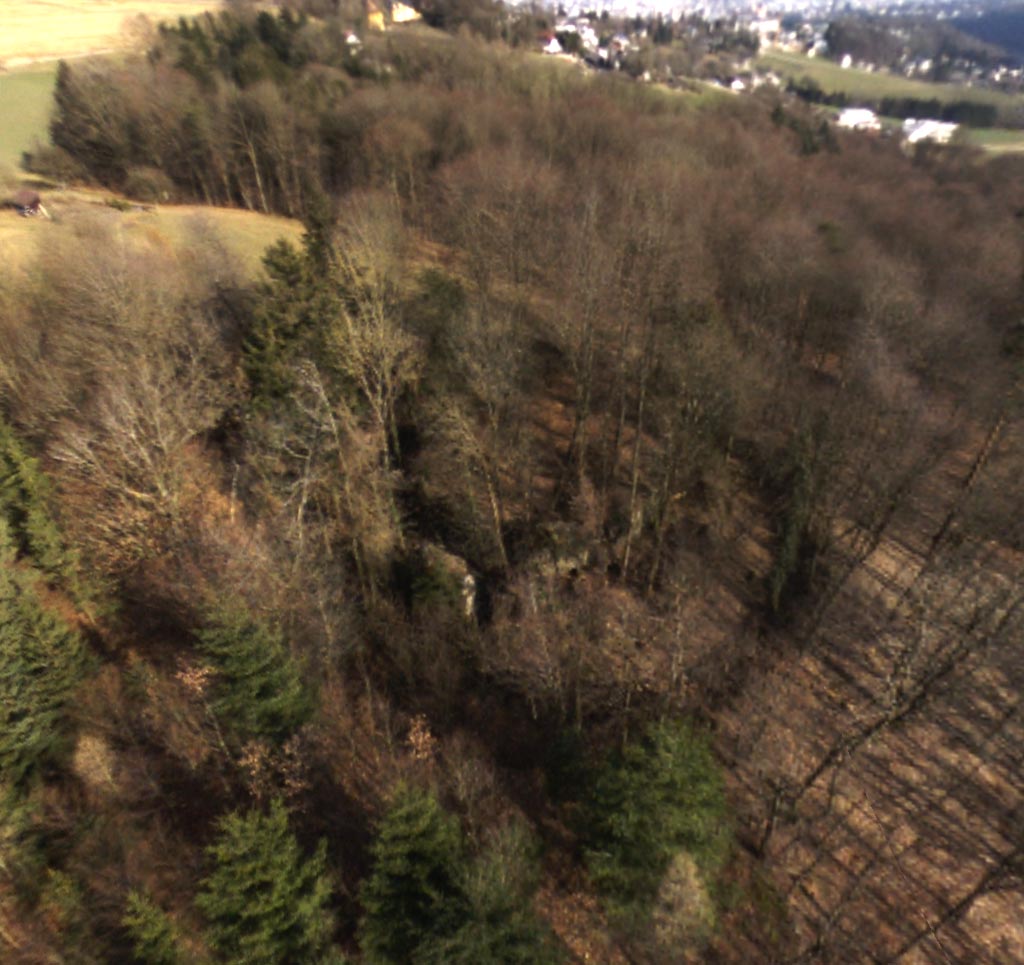}}
}%
\subfloat[][]{ 
	{\includegraphics[width=0.305\linewidth]{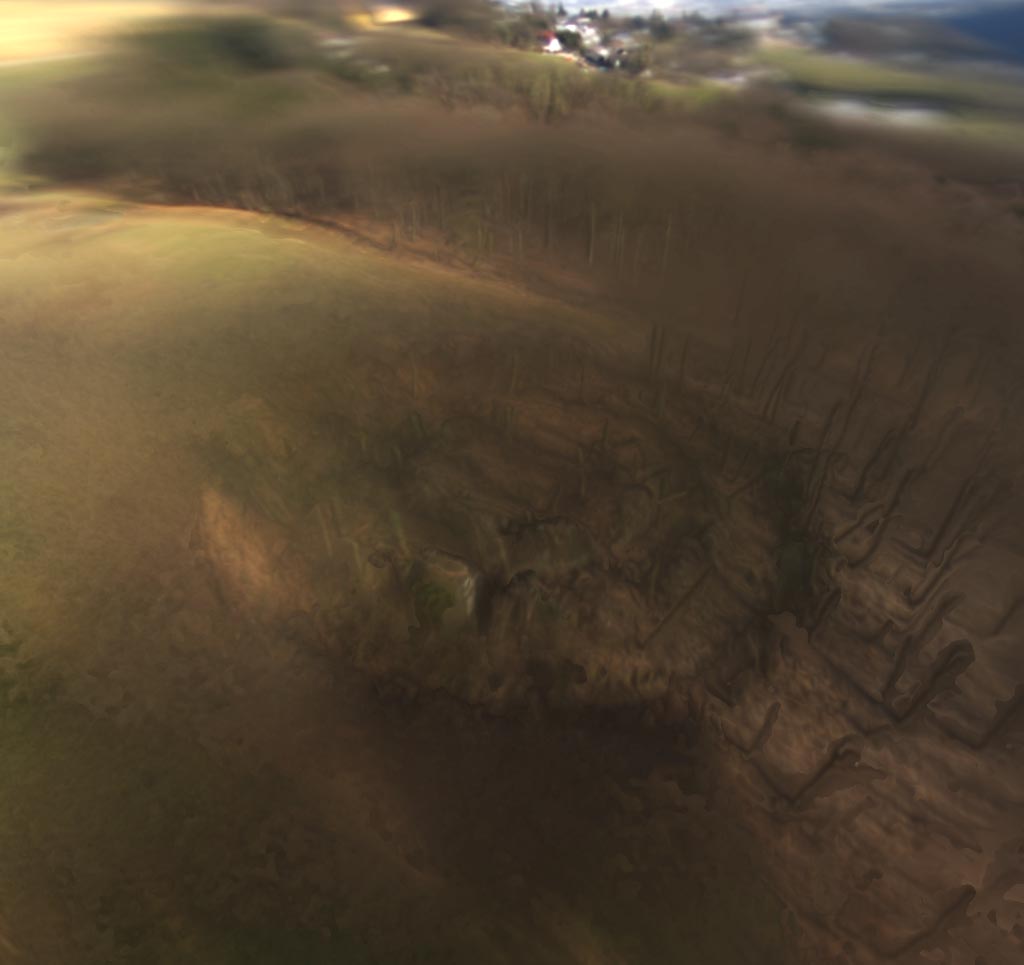}}
}%
	\caption{ A \SI{100}{m} diameter wide synthetic aperture (\SI{90}{\degree} sector only) sampled as an unstructured light field that consists of \num{504} drone images (a). %
	A single drone image does not reveal occluded artefacts on the ground (b). %
	With AOS the structure of remaining inner and outer ring walls and a trench become visible after optical sectioning (c). Note that, (a) shows satellite images used for path planning, while (b) and (c) show drone recordings. The circle in (a) illustrates the position of the ring walls revealed in (c).}
	\label{fig:aos}
\end{figure*}

With airborne optical sectioning (AOS) \cite{Kurmi2018,Bimberi2019IEEECGA}, we apply camera drones for synthetic aperture imaging. 
It samples the optical signal of wide synthetic apertures (up to \SI{100}{\m} diameter) with multiscopic video images as an unstructured (irregularly sampled) light field to support optical slicing by image integration. By computationally removing occluding vegetation or trees when inspecting the ground surface, AOS supports various applications in archaeology, forestry, agriculture, and border control. \autoref{fig:aos} illustrates an example where AOS was used to uncover the ruins of a 19th century fortification system that is concealed by dense forest and shrubs. The interested reader is referred to \cite{Kurmi2018,Bimberi2019IEEECGA} for more details.

Compared to alternative airborne scanning technologies (such as LiDAR) AOS is cheaper, delivers surface colour information, achieves higher sampling resolutions, and (in contrast to photogrammetry) it does not suffer from inaccurate correspondence matches and long processing times. However, as SART and SAI, AOS is passive (it only receives reflected light and does not emit electromagnetic or sound waves to measure the backscattered signal). Thus, it relies on an external energy source (i.e., sunlight). 
\\
\\
Common sense in synthetic aperture sampling is that a dense sampling pattern within a wide aperture is required to reconstruct a clear signal. In case of SAI, this implies that volumes of dense occluders require an unrealistically high number of image samples captured over a physically impractical aperture range. %
The disadvantage of a wide synthetic aperture for SAI is an increase of occlusion density for images captured at far distances and oblique angles (most extreme at the periphery of very wide synthetic apertures). 
Furthermore, the spatial resolution of camera recordings decreases with an increasing distance from the target object.
For SAI, the individual resolutions of all recordings are averaged. Thus, the reconstructed spatial resolution drops with an increasing aperture diameter. %
The disadvantage of a high sampling rate is the high processing demand that, if too high, prevents from real-time visualization rates. A wide synthetic aperture and a high sampling rate also increase the capturing time (if sampled sequentially, as for AOS) or the complexity of the sensor (if sampled simultaneously, as for camera arrays).
\\
\\
In this article we show that there exists practical limits to both, synthetic aperture size and number of samples. This leads to an understanding on how to design synthetic aperture sampling patterns and sensors in a most optimal and practically efficient way. 

We present three basic findings in this article: \textbf{(1)} There exists a limit to the baseline (distance) of sample positions. The minimal (optimal) baseline is the one that results in a disparity equal to the projected occluder size. Larger baselines do not improve visibility. \textbf{(2)} There exists a limit to achievable visibility improvement that depends on the density of the occluder volume. The maximum visibility gain is achieved at a density of \SI{50}{\percent}. \textbf{(3)} The normalized visibility gain (normalized to the density-dependent, achievable range) is independent of the occlusion density. It is directly correlated to a fixed number of samples. 

In the following, we will discuss these findings and explain how they lead to minimal synthetic apertures with a lowest number of samples. We present results based on a simplified mathematical model, simulations with a 3D LiDAR scan of forest (providing ground truth), and real AOS recordings of forest (without ground truth). For AOS recordings, we achieve an approximately \numrange{6.5}{12.75} times reduction of the synthetic aperture area and an approximately \numrange{10}{20} times lower number of samples without significant loss in visibility.
 
Although we focus on SAI (in particular AOS), our findings might be transferable to other SA sensors that support occlusion removal.  
\section{A Statistical Model for SAI}\label{sec:model}
The key idea of our mathematical model is to consider the sampling process of SAI not in the common context of optics, where a wide aperture leads to a shallow depth of field and large point-spread of out-of-focus occluders. Instead, we understand SAI sampling and reconstruction as a variation of signal averaging that is explainable by statistical principles. In fact, synthetic aperture rendering (the computational process to reconstruct new images from captured SAI samples) is nothing more than averaging images at proper disparity shifts that correspond to the adjusted synthetic focal plane \cite{Levoy1996,Isaksen2000}. 

If we consider the projected pixel-footprints of occluders as noise and the projected pixel-footprints of the target at a given synthetic focal plane as signal, we attenuate the occluders while amplifying the target when combining all SAI sample images. The reason for this is that all points on the selected synthetic focal plane that are imaged in all samples always project exactly to the same positions in the reconstructed image while all other points will project to different positions.
From signal averaging theory we know that the signal-to-noise ratio (S/N) of noisy measurements improves proportional to $\sqrt{\N}$ if $\N$ samples with constant signal are averaged \cite{SignalAveraging2010}. 
Thus, random noise in images is reduced by averaging multiple recordings of the same content. 
This principle does not apply to SAI as the random noise pattern (projected occluders) is mainly constant in each sample but differs only in disparity shifts that depend on sampling baselines and distances. 
 
\setlength{\fboxsep}{0pt}
\begin{figure}[b]
\centering
\subfloat[][]{ 
 {\scriptsize
	\def\svgwidth{0.29\columnwidth}
	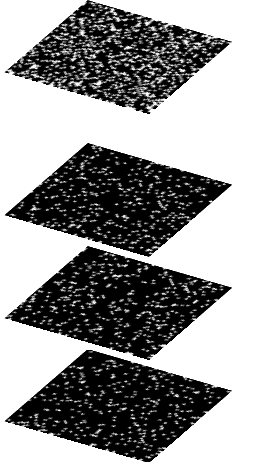}
}
\subfloat[][]{ 
	{\includegraphics[width=0.68\columnwidth]{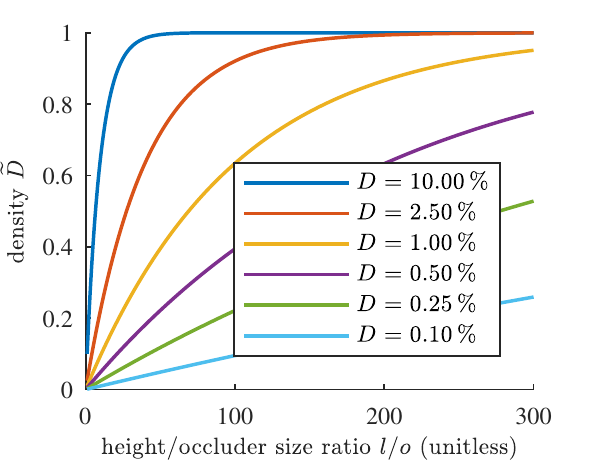}}
}
\caption{Our model assumes an occluder volume of height $\volumeHeight$ that consists of multiple slices with binary, uniformly distributed occluders of uniform size $\occluder$, and uniform density per slice of $\densityPerSlice$. After orthographically projecting all slices into an image, this results in an integrated density $\density$ for each SAI sample (a). The increase of integrated density $\density$ with respect to per-slice-density $\densityPerSlice$ is illustrated for different  volume height to occluder size  ratios $\volumeHeight/\occluder$ in (b), and is given by \ref{eq:integrated_density}.
}
\label{fig:model}
\end{figure}
Our model assumes a volume of height $l$, containing binary random occluders (assuming opaque occluders) of uniform size $o$ and cubic shape, uniform distribution (occluders cannot overlap in space), and uniform density $D$ at each slice of the volume (cf. \autoref{fig:model}a). Furthermore, it considers an orthographic projection for integrating all slices into sample images.

An occlusion density per slice of $\densityPerSlice$, an occluder size of $\occluder$, and a volume height of $l$ leads to an integrated occlusion density for each SAI sample after orthographic projection of  

\begin{equation} \label{eq:integrated_density}
\density = 1 - (1-\densityPerSlice)^{\volumeHeight/\occluder}, \end{equation}

where the ratio $\volumeHeight/\occluder$ is unitless. A light ray passing through the volume in our model follows a binomial distribution. In \autoref{eq:integrated_density} $(1-\densityPerSlice)^{\volumeHeight/\occluder}$ is the probability mass function of a non-occluded ray (interacting with zero occluders within its path through the volume). Thus, statistically, $\density$ is the probability of occlusion in a single SAI sample (cf. \autoref{fig:model}b).
\begin{figure*}%
	\centering%
	\subfloat[][]{ 
		{\includegraphics[width=0.32\linewidth]{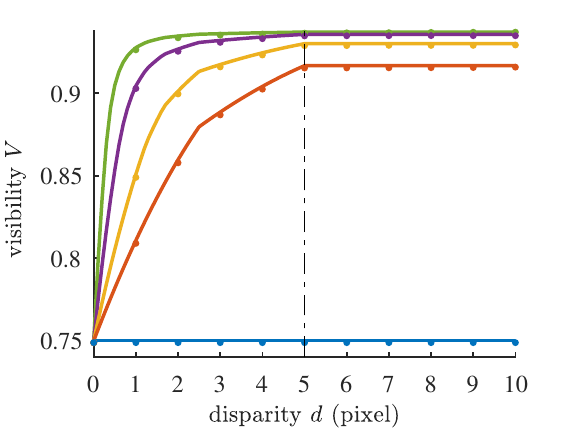}}
	}%
	\subfloat[][]{ 
		{\includegraphics[width=0.32\linewidth]{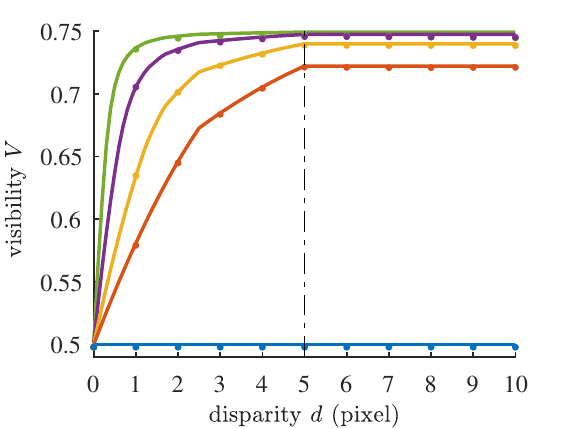}}
	}%
	\subfloat[][]{ 
		{\includegraphics[width=0.32\linewidth]{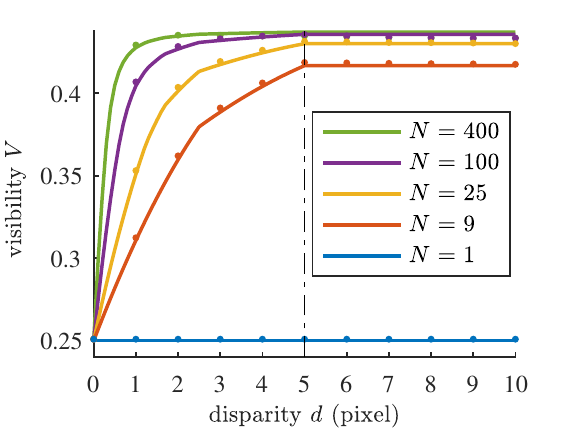}}
	}%
	\caption{Visibility improvements achieved with varying disparity $\disparity$, different SAI sample densities: $\density = \SI{25}{\percent}$ (a), $\density = \SI{50}{\percent}$ (b), and $\density = \SI{75}{\percent}$ (c), and an increasing number of SAI samples $N$. %
		The projected occluder size is $\occluder = 5$ pixels in all cases. The dot-dashed lines indicate the optimal disparity, where $\nd = \disparity / \occluder = 1$. %
		While solid lines are results computed from \ref{eq:visbility}, points represent simulation results (averaged noise patterns as in \autoref{fig:simImg}).}
	\label{fig:sim_vis_d}
\end{figure*}
When averaging $N$ such SAI samples that are 2D shifted by disparity $d$, we statistically achieve a visibility probability of

\begin{equation} \label{eq:visbility}
\begin{split}
\visibility = & 1 - \p^2 - \frac{\p(1-\p)}{\N^2} \times \\%
		& \Bigg(\N +4 \sum_{i=1}^{\sqrt{\N}-1}\sum_{j=0}^{\sqrt{\N}-1}(\sqrt{\N}-i)(\sqrt{\N}-j) \\%
	  & \Max(0, (1-i\nd))\Max(0, (1-j\nd))\Bigg),
\end{split}
\end{equation}

with $\nd = \disparity / \occluder$ being the ratio of projected occluder size $o$ and disparity shift $d$. Note that in \ref{eq:visbility}, $o$ and $d$ are in SAI sample units (projected pixel distances in camera resolution), but $\nd$ is unitless. The derivation of \ref{eq:visbility} is provided in the Appendix.

\autoref{fig:sim_vis_d} illustrates the visibility improvement over varying disparity choices, different SAI sample densities, and an increasing number of SAI samples. In all cases, the visibility improvement settles at a constant maximum after an optimal disparity. This optimal disparity equals the projected occluder size $o$, and is reached when $\nd=1$. 

\begin{figure*}[t]%
	\centering%
	\includegraphics[width=\linewidth]{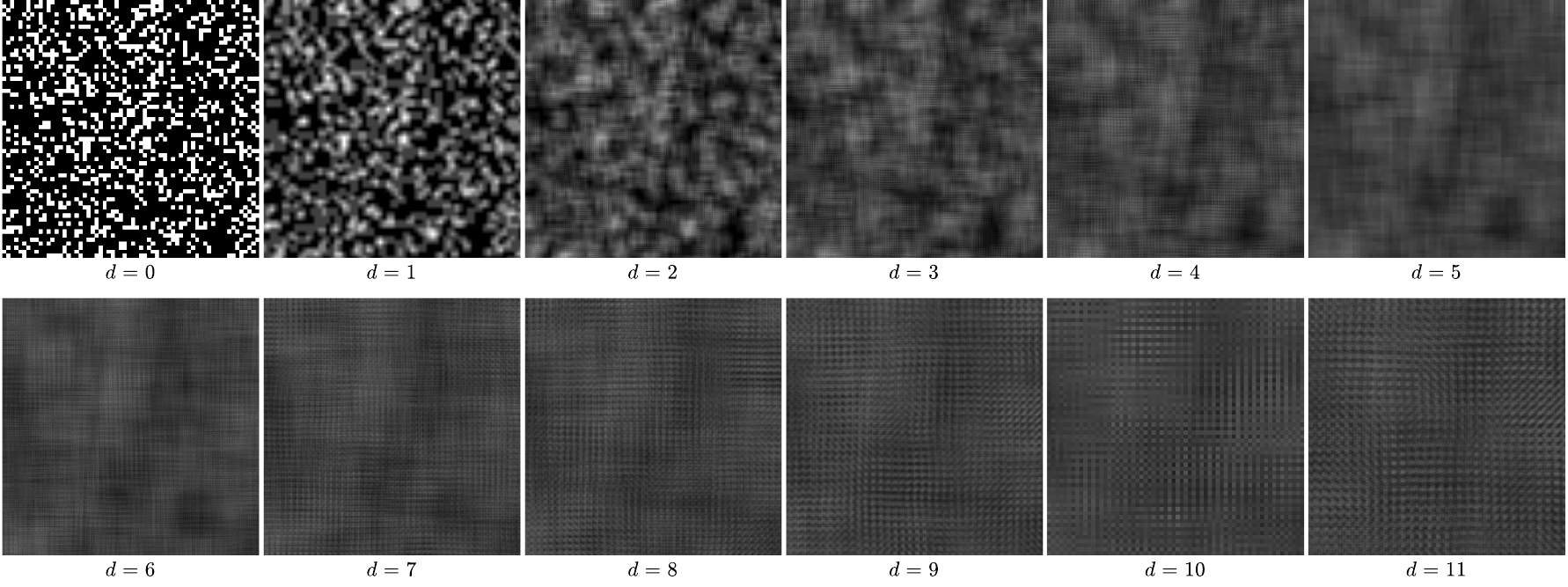}
	\caption{Visual simulation results for a random binary noise pattern ($\density = \SI{25}{\percent}$, $\occluder=5$) being disparity shifted $\N=100$ times before averaging. 
	The noise pattern is shifted in $10 \times 10$ two-dimensional steps of step width $d$.
	Noise reduction settles at $d=5$ ($\nd=1$). With $d>5$ ($\nd>1$) only bokeh artefacts that visualize the sampling grid are emphasised. Visibility values for this example are indicated by the blue dots (measured in the resulting images) / line (computed with our model) in \autoref{fig:sim_vis_d}a.}
	\label{fig:simImg}
\end{figure*}

\autoref{fig:simImg} shows visual results of a simulation. Noise reduction (visibility improvement with respect to the black background) settles at $d=5$ ($\nd=1$). With $d>5$ ($\nd>1$) only bokeh artefacts that visualize the sampling grid are emphasized. In \autoref{fig:sim_vis_d}, it is also shown that all such simulations match our model \ref{eq:visbility} precisely and under all conditions.

If physical capturing conditions, such as intrinsic camera parameters, resolution, and recording distance (flying altitude in case of AOS) are known, disparity in pixel directly translates to camera baseline in meters. This explains our first finding: \textbf{(1)} There exists a limit to the baseline (distance) of sample positions. The minimal (optimal) baseline is the one that results in a disparity equal to the projected occluder size. Larger baselines do not improve visibility. 

If the optimal (or larger) disparity is achieved, then \ref{eq:visbility} can be simplified for $\nd  \geq 1$ to

\begin{equation} \label{eq:visbility_simplified}
\visibility = 1-\p^2 - \frac{\p(1-\p)}{\N}.
\end{equation}

\begin{figure*}[b]%
	\centering%
	\subfloat[][]{ 
	{\includegraphics[width=111mm]{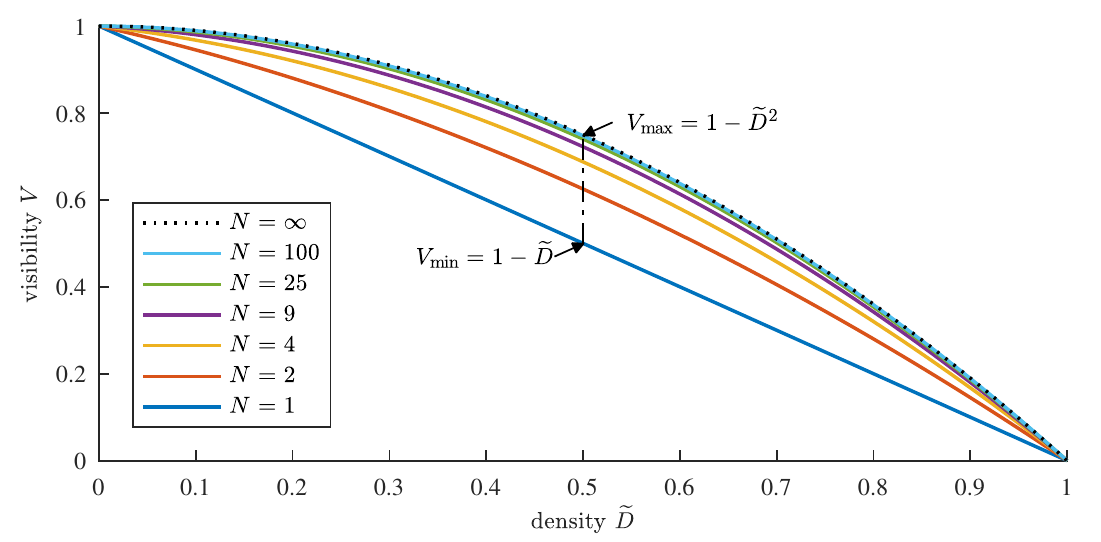}}
}%
\subfloat[][]{ 
	{\includegraphics[width=69mm]{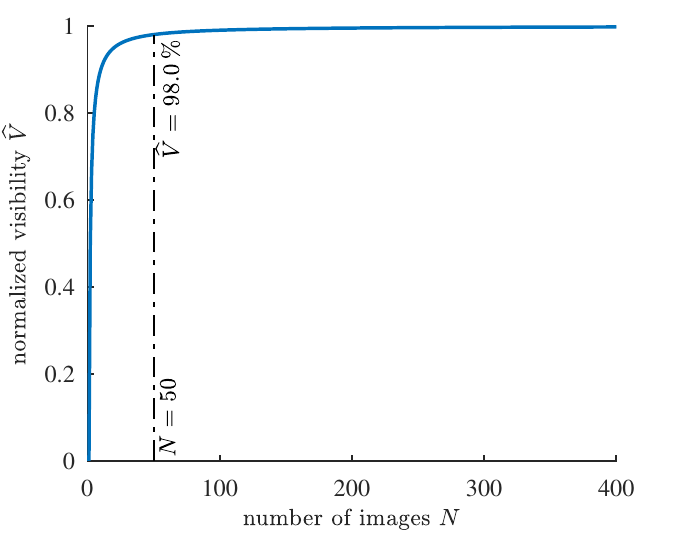}}
}%
	\caption{Our model \autoref{eq:visbility_simplified} with $\nd  \geq 1$ for varying $\density$ and $\N$ (a). The possible visibility gain is always in a range between $1-\density$ and $1-\density^{2}$, and  maximizes at $\density=\SI{50}{\percent}$. The normalized visibility $\normVisibility$ (normalized to $V_\text{min}$--$V_\text{max}$ range) becomes independent of the density and settles quickly at a low $N$ (b). Picking a tolerable $\normVisibility$-threshold of, for instance \SI{98}{\percent}, results in constant $N$ of \num{50} SAI samples that have to be captured and averaged---regardless of the occlusion density.}
	\label{fig:calc_vis_D}
\end{figure*}

\autoref{fig:calc_vis_D}a plots \ref{eq:visbility_simplified} over various $\density$ and $N$, and illustrates that there exists a lower bound of $V_\text{min}=1-\density$ for $N=1$ and an upper bound of $V_\text{max}=1-\density^{2}$ for $N=\infty$. It also shows that the maximal visibility gain is achieved at $\density=\SI{50}{\percent}$.

This explains our second finding: \textbf{(2)} There exists a limit to achievable visibility improvement that depends on the density of the occluder volume. The maximum visibility gain is achieved at a density of \SI{50}{\percent}.

If we normalize \ref{eq:visbility_simplified} within the possible range ($V_\text{min}$ to $V_\text{max}$), it becomes independent of the density (cf. \autoref{fig:calc_vis_D}b):

\begin{equation} \label{eq:normVisibility} 
\normVisibility = ({\visibility}\!-\! V_\text{min})/(V_\text{max} \! - \! V_\text{min}) = 1 - \frac{1}{\N}.
\end{equation}

This explains our third finding: \textbf{(3)} The normalized visibility gain is independent of the occlusion density. It is directly correlated to a fixed number of samples. Thus, a predefined tolerable visibility threshold leads to a constant number of required SAI samples---regardless of the occlusion density.
\\
\\
Note, that in reality we can neither consider sizes, densities, and distributions of occluders to be uniform, nor that orthographic projection applies to regular cameras. However, the idealized model described above opens a statistical take on SAI for making sampling decisions. In the subsequent chapter we compare our model with realistic data and show that our findings still hold.

\section{SAI under realistic conditions}\label{sec:scan}
\begin{figure*}
	\centering%
	{ \scriptsize %
		\def\svgwidth{0.99\linewidth} %
		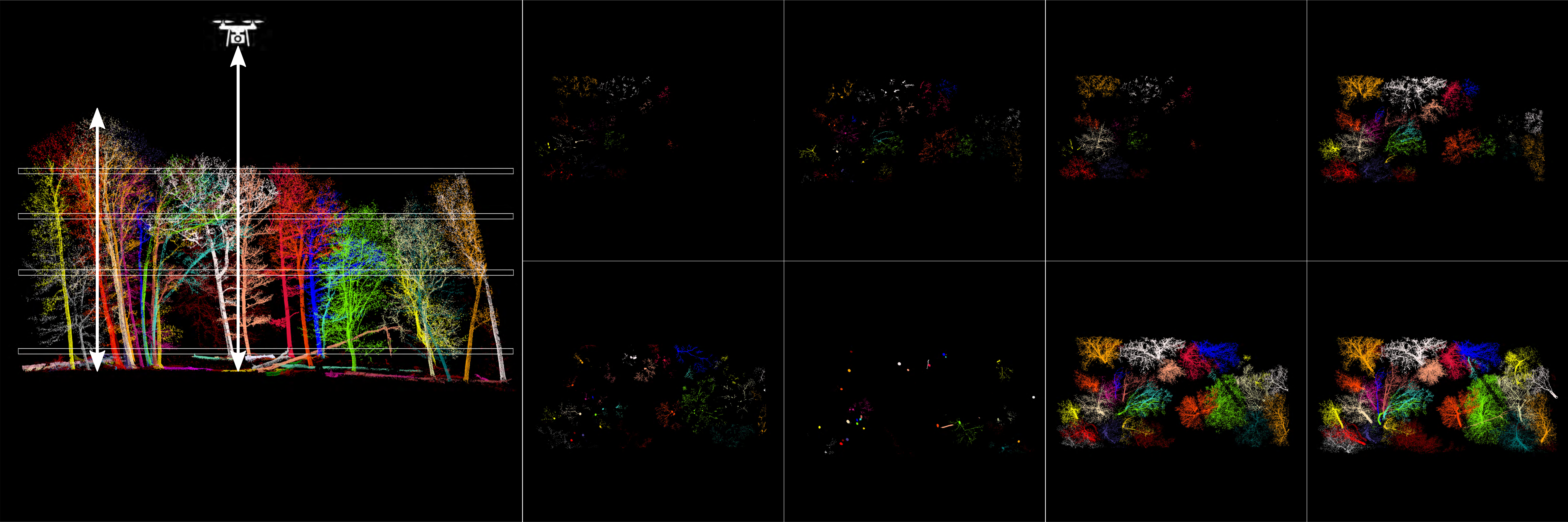%
	}
	\caption{A LiDAR scan of a forest patch results in a volume of 3D points. Choosing four representative slices throughout the volume gives an indication on what realistic average values for per-slice density $\densityPerSlice$, and occluder size $o$ are. %
Integrated densities $\density$ over various ranges (from the top to corresponding slices) are shown on the right. The simulation at the bottom right corner (top -- $4$; $\density=\SI{53.7}{\percent}$) represents a single SAI sample captured with perspective projection from an altitude of \SI{50}{\meter} from the synthetic focal plane.}
	\label{fig:leaftreemodel}
\end{figure*}
%
We can compare our model with the non-uniform occlusion volume of a forest patch that has been scanned using LiDAR \cite{Trochta2017} (\autoref{fig:leaftreemodel}).

\begin{figure*}[b]
	\centering%
	\subfloat[][]{ 
	\includegraphics[width=\columnwidth]{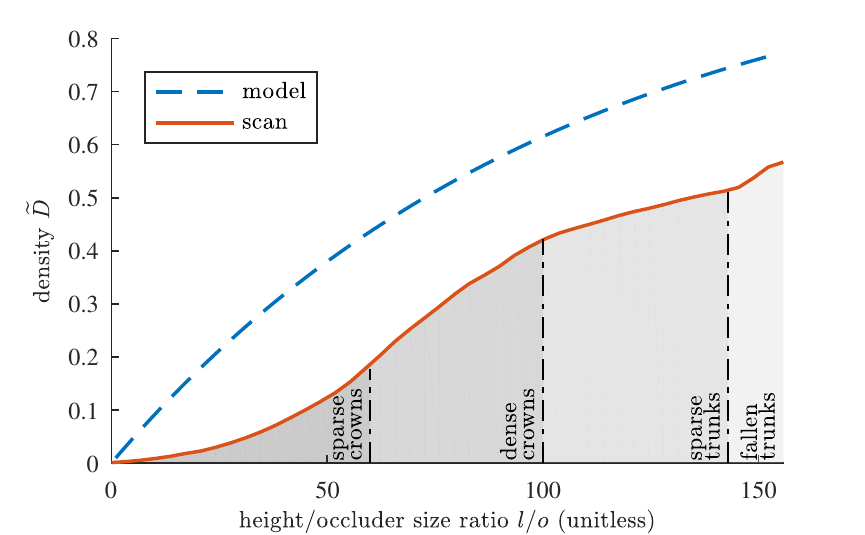}
	}
	\subfloat[][]{ 
	\includegraphics[width=\columnwidth]{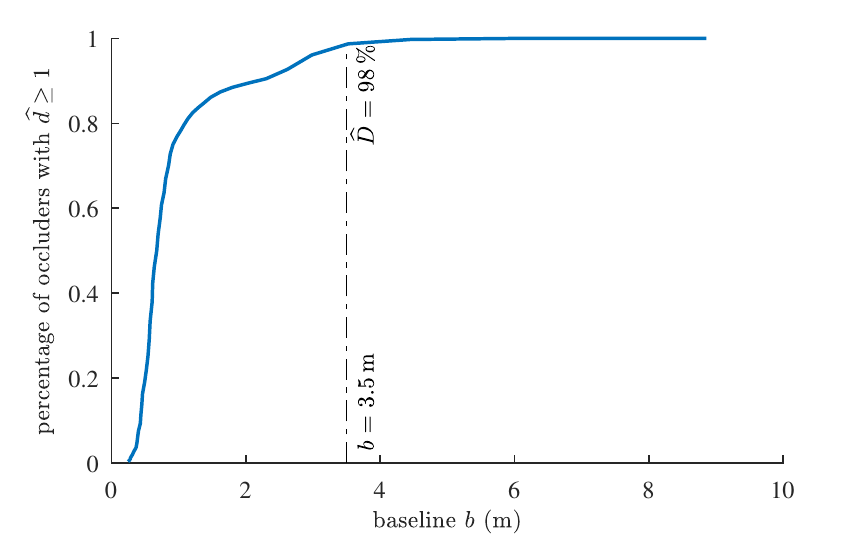}
	}
	\caption{Integrated density $\density$ predicted by our model with a per-slice density of $\densityPerSlice = \SI{0.95}{\percent}$ (average of the entire forest volume, \autoref{fig:leaftreemodel}) and directly computed from the LiDAR scan (a). Percentage of occluders that project the optimal disparity ($ \normDisp \geq 1$) when the baseline $b$ between adjacent SAI samples (drone positions in AOS) is varied (b).}
	\label{fig:scan}
\end{figure*}

\autoref{fig:scan}a plots $\density$ over $l/o$ for the LiDAR scan and for our model \ref{eq:integrated_density} using average values  that are determined through the entire volume. The non-uniformity (occluder sizes, densities, disparities) of the LiDAR scan as well as the different projections (orthographic projection in the model vs. perspective projection in the scan) leads to the deviation from our model prediction. However, the tendential behaviour is the same. Furthermore, our model represents a worst-case upper bound in $\density$ since it considers uncorrelated noise while, in reality, individual occluder points are correlated (they form connected segments, such as branches and trunks).   

\autoref{fig:scan}b plots for the LiDAR scan the percentage of occluders whose sizes are projected to the optimal disparity ($ \normDisp \geq 1$) over different choices of baselines $b$. It indicates, for example, that from an axial sampling altitude of \SI{50}{\meter} and for a baseline (lateral sampling distance) of \SI{3.5}{\meter}, \SI{98}{\percent} of all occluders project to equal or smaller pixel footprints than their corresponding disparities achieved with the baseline.   

The size of the synthetic aperture (the lateral scanning area) depends on the chosen scanning baseline $b$ that directly relates to  a particular $\normDisp$, and on the number of SAI samples $N$ being captured: 

\begin{equation} \label{eq:aperture} 
\aperture =  (\sqrt{\N}-1) b,
\end{equation}

where $a$ is the aperture diameter. Thus, multiple combinations of $N$ and $b$ (corresponds to $\normDisp$) lead to the same $a$. Our goal is to find a combination that minimizes $a$ (to maximize spatial resolution in the reconstruction and to reduce occlusion density due to oblique viewing angles) and $N$ (to support real-time visualization rates and minimize sampling time) that leads to a tolerable visibility gain.  

\begin{figure*}
	\centering%
	\subfloat[][]{ 
	\includegraphics[width=\columnwidth]{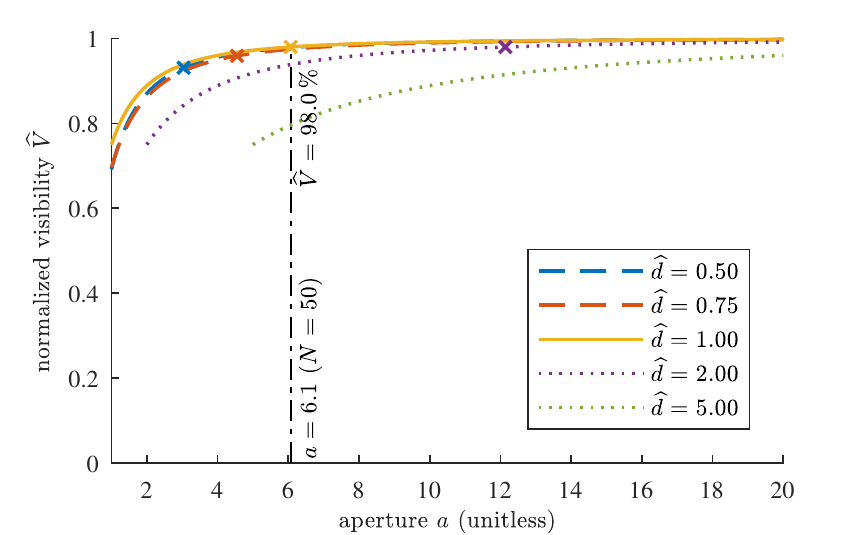}
	}
	\subfloat[][]{ 
	\includegraphics[width=\columnwidth]{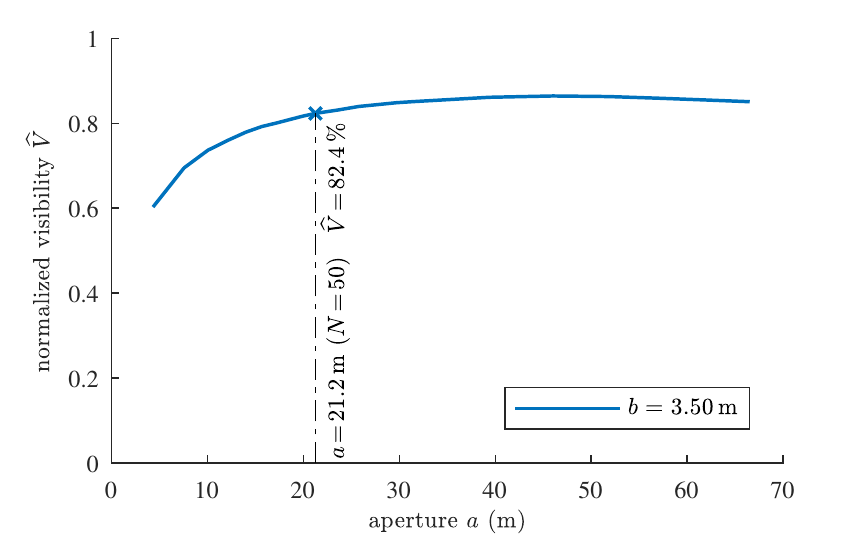}
	}
	\caption{Normalized visibility $\normVisibility$ for different aperture sizes $\aperture$ and varying $\normDisp$ computed with our model (a). The dashed lines show oversampling causing too small disparities. The dotted lines indicate undersampling due to too large baselines. The x-markers indicate when $\N=50$ samples are required within a given $\aperture$ to achieve a particular $\normDisp$. Normalized visibility of the LiDAR scan with respect to the aperture size and for a pre-selected baselines of $b=\SI{3.5}{\m}$ (b). }
	\label{fig:aperture}
\end{figure*}
When plotting the normalized visibility $\normVisibility$ with respect to aperture diameter that would be needed to achieve a particular $\normDisp$ with our model (\autoref{fig:aperture}a), it can be seen that for the optimal $\normDisp=1$ and for a chosen visibility threshold of $\normVisibility=\SI{98}{\percent}$ (which correlates to a fixed $N$ of \num{50} samples \ref{eq:normVisibility}), the required aperture diameter is \num{6.1} (unitless in our model). But \autoref{fig:aperture}a also shows that if $\normDisp$ is lower than \num{1} the maximal visibility cannot be achieved for smaller apertures because projected occluder footprints are larger than the achieved disparities. In this case, the synthetic aperture is oversampled. If $\normDisp$ is greater than \num{1}, the maximal visibility can also not be achieved for smaller apertures because of undersampling (the corresponding baselines are too large to support an adequate $N$ within the aperture).  
The same plot is shown for the LiDAR scan in \autoref{fig:aperture}b, using the previously chosen baseline of $b=\SI{3.5}{\meter}$ and $N=50$ which (based on our mode) should lead to $\normVisibility=\SI{98}{\percent}$. The reason why it leads only to $\normVisibility=\SI{82.4}{\percent}$ is the non-uniformity of the forest that does not perfectly match our uniform model, as explained in \autoref{sec:model}. 
\begin{figure*}
	\centering%
 \ensuremath{
	\scriptsize
	\def\svgwidth{0.905\linewidth}
	\input{figures/LeafTreeScan_ImageMatrix_N_vs_d.pdf_tex}}
	\caption{Visual comparison of synthetic aperture rendering results for varying $\baseline$ and $\N$ for the LiDAR scan (green) overlaid over a grayscale EIA resolution chart as target. The synthetic focal plane is aligned with the target plane. SSIM values are the structural similarity index \cite{wang2004image} when comparing against the theoretical optimum ($N=\infty$).}
\label{fig:scanmatrix}
\end{figure*}

 \autoref{fig:scanmatrix} illustrates visual synthetic aperture rendering results of the LiDAR scan (green) overlaid over a grayscale EIA resolution chart for increasing $N$ and $b$. The synthetic focal plane is located at the target plane, and the axial sampling distance is \SI{50}{\m} (\autoref{fig:leaftreemodel}). The visibility $\normVisibility$ is computed based on the fraction of non-occluded parts on the target plane. Note, that the theoretical optimum of SAI (for $\N=\infty$; $\normVisibility=\SI{100}{\percent}$) is the focussed image of the target at the synthetic focal plane overlaid by a constant bias which corresponds to the mean occlusion density ($\p$). For the LiDAR scan, this is $\p=\SI{56.7}{\percent}$, and the theoretical optimum will be an image of the EIA resolution chart with a \SI{56.7}{\percent} greenish tint in our simulation. We use the structural similarity index (SSIM) \cite{wang2004image} as a quantitative measure of visual difference between this theoretical optimum and other visual synthetic aperture rendering results. 
The SSIM is a common image quality metric which considers the fact that humans are highly adapted to extract structural information.
 \autoref{fig:scanmatrix} illustrates that SSIM values and visibility values are correlated, and that for baselines or number of SAI samples that differ from our selection ($b=\SI{3.5}{\m}$, $N=50$), no significant improvement is achieved (only \SIrange{4}{5}{\percent} in visibility and SSIM). 
\\
\\
%
\begin{figure*}
	\centering%
 \ensuremath{
	\scriptsize
	\def\svgwidth{0.99\linewidth}
	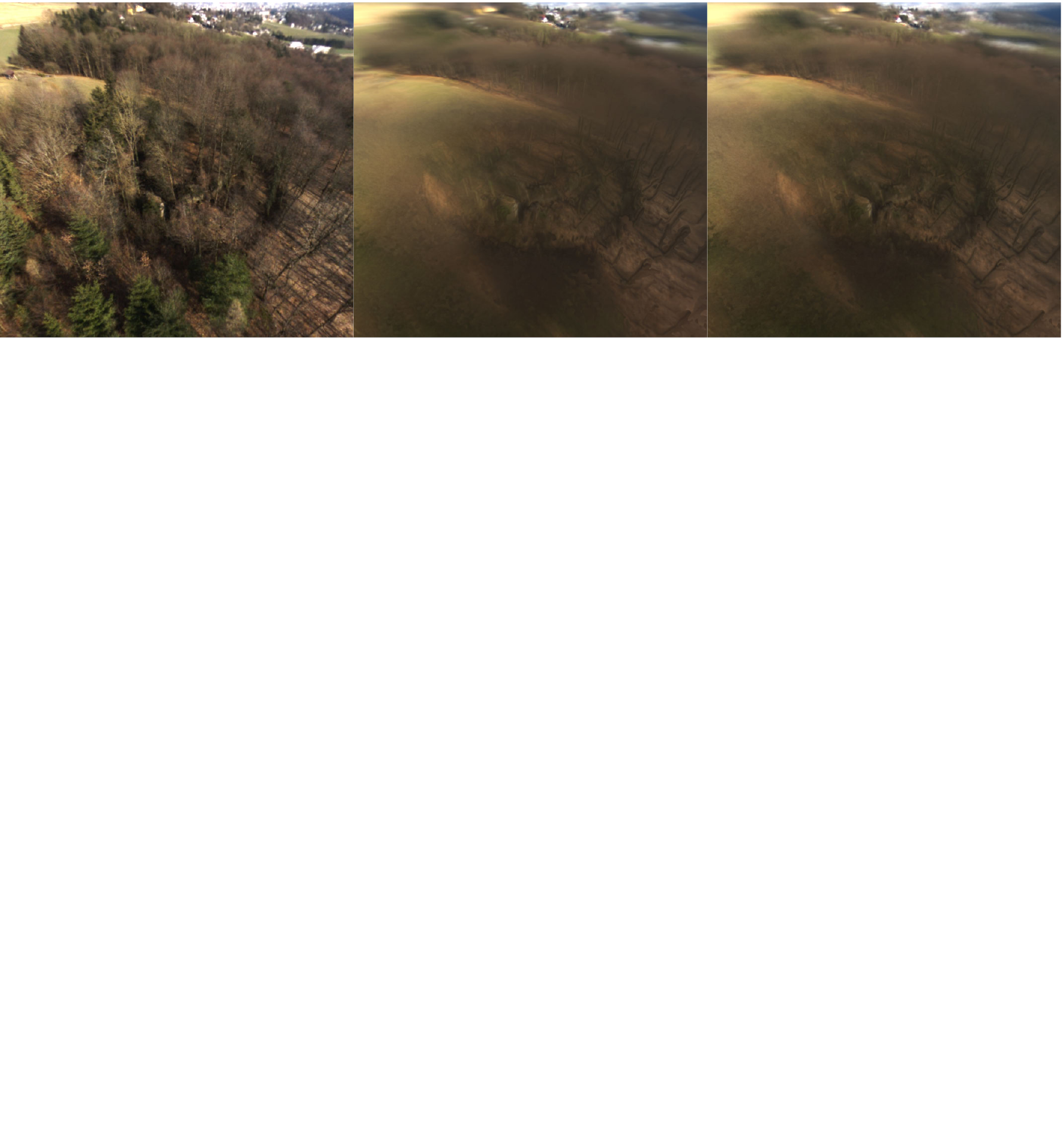}
	\caption{Synthetic aperture rendering results of the AOS scan presented in \cite{Kurmi2018} for different $\aperture,\N$ combinations that satisfy the baseline suggested by our model ($\baseline = \SI{2.8}{\meter}$). Visual difference to the fully sampled case ($a=\SI{50}{\meter}$, $N=\num{504}$) is indicated by SSIM \cite{wang2004image}.}
	\label{fig:tower16}
\end{figure*}
%
We can now apply our findings to the real AOS scan shown in \autoref{fig:aos} \cite{Kurmi2018}: This scan was initially brute force sampled from an axial distance (altitude) of $\SI{40}{\m}$, with a baseline of $b=\SI{2}{\m}$, and with $N=504$ samples within a synthetic aperture of $\SI{2500}{\square\meter}$ ($a=\SI{50}{\m}$). If we consider the forest in the LiDAR dataset to be statistically representative to the forest in the AOS scan, then our model would suggest a baseline of $b=\SI{2.8}{\m}$ (with respect to the different axial scanning distances in both cases:  $b=\SI{3.5}{\m} \cdot \SI{40}{\m}/\SI{50}{\m}$). This is the baseline at which we expect $\SI{98}{\percent}$ of all occluders to project to the optimal disparity ($ \normDisp \geq 1$), as shown in \autoref{fig:scan}.

\autoref{fig:tower16} illustrates synthetic aperture rendering results of the AOS scan at a varying numbers of SAI samples $N$ that, together with the fixed baseline of $b=\SI{2.8}{\m}$ lead to particular synthetic apertures of diameters $a$.  With $N=50$ samples, we expect (based on the chosen normalized visibility threshold of $\normVisibility=\SI{98}{\percent}$, as shown in \autoref{fig:aperture}a and in \autoref{fig:calc_vis_D}b) no significant gain in visibility for a larger number of SAI samples. But we are aware of the fact that the  visibility which can be practically achieved will be lower than $\SI{98}{\percent}$ due to the non-uniformity of the forest, as shown in \autoref{fig:aperture}b. Since for the AOS scan no ground truth exists, we cannot determine visibility in a quantitative way. Instead, we compare new sampling results against the brute-force sampled result (using $N=504$, $a=\SI{50}{m}$, $b=\SI{2}{m}$) that we consider an approximation to the theoretical optimum. Visual similarity is determined again with the structural similarity index metric (SSIM) \cite{wang2004image}.  

\begin{figure}[b!]
	\centering%
	\includegraphics[width=\columnwidth]{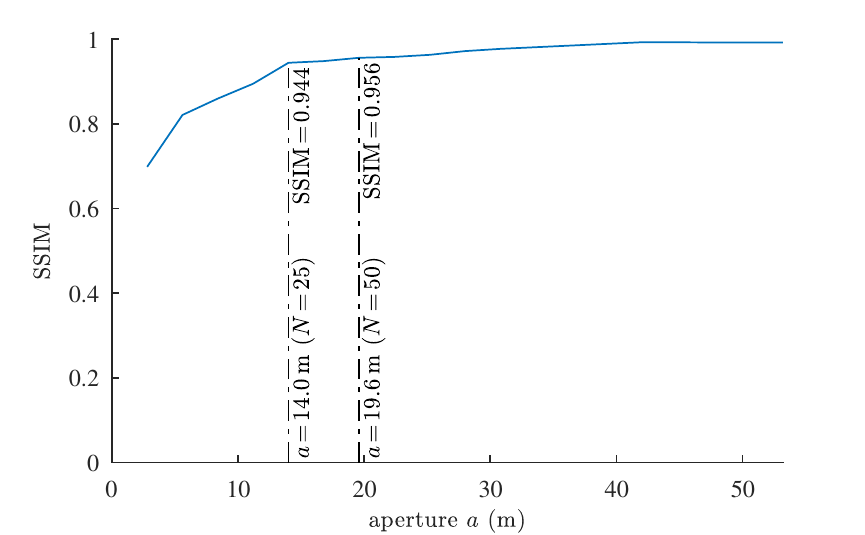}
	\caption{SSIM values for synthetic aperture rendering results of the AOS scan with varying aperture diameters $a$ and fixed baseline $\baseline = \SI{2.8}{\m}$. The reference is the fully sampled result with $a=\SI{50}{\m}$ and $N=504$.}
	\label{fig:tower16_ssim}
\end{figure}

\autoref{fig:tower16_ssim} plots all SSIM values for $a \leq \SI{50}{\m}$, and indicates that even an aperture diameter of $a=\SI{14}{\m}$ with $N=25$ SAI samples (this corresponds to a normalized visibility threshold of $\normVisibility=96\%$) would lead to no significant visual reduction compared to an aperture of $a=\SI{50}{\m}$ with $N=504$ samples.
%
%
These results imply that the synthetic aperture for this scene can be significantly smaller (by a factor of $\approx$ \numrange{6.5}{12.75} in area and by $\approx$ \numrange{10}{20} in number of samples).

\section{Conclusion and Future Work}
In this article we have presented three findings that lead to a basic understanding on how to design synthetic aperture sampling patterns: \textbf{(1)} There exists a limit to the baseline (distance) of sample positions. The minimal (optimal) baseline is the one that results in a disparity equal to the projected occluder size. Larger baselines do not improve visibility. \textbf{(2)} There exists a limit to achievable visibility improvement that depends on the density of the occluder volume. The maximum visibility gain is achieved at a density of \SI{50}{\percent}. \textbf{(3)} The normalized visibility gain is independent of the occlusion density. It is directly correlated to a fixed number of samples. 
For AOS and the evaluated datasets, these findings result in much smaller synthetic aperture areas with significantly less samples.

The key idea of our approach is to consider the sampling process of SAI not in the common context of optics, where a wide aperture leads to a shallow depth of field and large point-spread of out-of-focus occluders. Instead, we understand SAI sampling and reconstruction as a variation of signal averaging that is explainable by statistical principles.

In future, we want to explore better models that consider the non-uniformity of known occlusion volumes, such as forests where (from tree crowns to the ground) occluder size increases while density decreases. 
Furthermore, we want to investigate the potential non-uniform coded sampling patterns, and techniques that enable a measurement or a better approximation of local occluder sizes and densities. 
Finally, we are interested in evaluating our model with different forest types (such as conifer or rain forest) and datasets. Thereby, measurements and records from forestry research might guide the parametrization of occluder sizes and densities.

\appendix[Derivation of the Visibility Probability]
For a given focal plane, synthetic aperture rendering integrates $N$ disparity-shifted SAI samples $Y_i$:   
\begin{equation}\label{eq:app_x_sum}
X =  \frac{1}{\N}\sum_{i=1}^{\N}Y_i.
\end{equation}

We define visibility $\visibility$ based on the mean squared error (MSE) between the synthetic aperture rendering result $X$ and an  image $R$ of the non-occluded target on the focal plane:
\begin{equation}\label{eq:visibility}
\visibility = 1 - \MSE = 1 -  \E[{\big(X - R\big)}^{2}],
\end{equation}

where $\E$ indicates the expectation (i.e., mean). For simplicity, we assume $R = 0$ in the following, hence
\begin{equation}\label{eq:app_mse_x}
\MSE  =  \E[X^{2}] = \E[X]^2 + \Var[X].
\end{equation}

Each SAI sample has equal mean and variance (i.e., $\E[Y_i] = \E[Y_k]$ and $\Var[Y_i] = \Var[Y_k]$), behaves like a Bernoulli trial, and is a mixture of Bernoulli distributions:
\begin{equation}\label{eq:app_mean}
\E[X] = \E \bigg[ \frac{1}{\N}\sum_{i=1}^{\N}Y_i \bigg] %
= \frac{1}{\N}\sum_{i=1}^{\N} \E \big[Y_i \big]%
= \E \big[ Y_{i} \big],%
\end{equation}

\begin{equation}\begin{split}\label{eq:app_var}
\Var[X] =& \Var \bigg[ \frac{1}{\N}\sum_{i=1}^{\N}Y_i \bigg]\\ %
=& \frac{1}{\N^2} \bigg( \sum_{i=1}^{\N} \Var[Y_i] + 2 \sum_{k > i} \Cov[ Y_i,Y_k ] \bigg) \\ %
=& \frac{1}{\N^2} \bigg( \N \Var[Y_{i}] + 2 \sum_{i=1}^{\N} \sum_{k > i} \Cov[ Y_i,Y_k ] \bigg),
\end{split}\end{equation}

\begin{equation}\label{eq:app_cov}
\Cov[ Y_i,Y_k ] = \E[ Y_iY_k ] - \E[Y_i] E[Y_k]. %
\end{equation}

The probability that the same occluder is not projected to the same pixel in two SAI samples ($Y_i$ and $Y_k$) is $\q_{ik} = \Min( 1,  \disparity_{ik}/ \occluder)$, where $\disparity_{ik}$ is the total disparity of the occluder in $Y_i$ and $Y_k$. 

Because of the conditional dependence between $Y_i$ and $Y_k$ (being shifted instances of the same image), $\E[ Y_iY_k ]$ can be expressed as

\begin{equation}\begin{split}\label{eq:app_exy} 
\E[ Y_iY_k ] =& \E \big[ \E[Y_iY_k \, | \,  Y_i] \big] %
= \E \big[ Y_i \E[Y_k \, | \,  Y_i] \big] \\ 
=& \E \big[ Y_i \big( (1-\q_{ik}) Y_i + \q_{ik} Y_k \big) \big] \\ %
=& (1-\q_{ik}) \E[ Y_i^2 ] + \q_{ik} \E[Y_i] \E[Y_k]  \\ %
=& (1-\q_{ik}) \big( \E[ Y_i ]^2 + \Var[Y_i] \big)+ \q_{ik} \E[Y_i] \E[Y_k]. %
\end{split}\end{equation}

Putting \ref{eq:app_exy} into \ref{eq:app_cov} yields
\begin{equation}\label{eq:app_cov_simple}
\begin{split}
\Cov[ Y_i,Y_k ] = & (1-\q_{ik}) \big( \E[ Y_i ]^2 + \Var[Y_i] \big) + \\ 
 & \q_{ik} \E[Y_i] \E[Y_k] - \E[Y_i] \E[Y_k]\\ %
= & (1-\q_{ik}) \big( \E[ Y_{i} ]^2 + \Var[Y_{i}] \big)+ \q_{ik} \E[Y_{i}]^2 - \E[Y_{i}]^2 \\
= & (1-\q_{ik}) \Var[Y_{i}].%
\end{split}\end{equation}

Thus, if $\q_{ik} = 1$ (when $\disparity_{ik}/o \geq 1$) the covariance $\Cov[ Y_i,Y_k ] = 0$.
Putting \ref{eq:app_cov_simple} into \ref{eq:app_var} and simplifying yields:
\begin{equation}
\begin{split}\label{eq:app_var_simple}
\Var[X] %
=& \frac{1}{\N^2} \bigg( \N \Var[Y_{i}] + 2 \sum_{i=1}^{\N} \sum_{k > i} (1-\q_{ik}) \Var[Y_{i}] \bigg) \\
=& \frac{\Var[Y_{i}]}{\N^2} \bigg( \N  + 2 \sum_{i=1}^{\N} (N-i) (\Max(  0, 1 - i \disparity / \occluder )) \bigg).
\end{split}\end{equation}

Note, that only the co-variance depends on $\disparity$ and $\occluder$. Putting \ref{eq:app_var_simple} and \ref{eq:app_mean} into \ref{eq:app_mse_x} yield: 

\begin{equation} \label{eq:app_mse_1D} 
\begin{split}
\E[X^2] = &  \E[Y_{i}]^2 + \frac{\Var[Y_{i}]}{\N^2} \times \\
& %
 \bigg( \N  + 2 \sum_{i=1}^{\N} (N-i) (\Max(  0, 1 - i \disparity / \occluder )) \bigg) \\
= & \p^2 + \frac{\p(1-\p)}{\N^2} \cdot \\%
	&\bigg(\N +2 \sum_{i=1}^{\N}(\N-i)\Max \big(  0, (1-i {\disparity}/{\occluder}) \big) \bigg).
\end{split}
\end{equation}

Note, that \ref{eq:app_mse_1D} considers 1D disparities, only. But it can be easily extended for 2D disparities:
\begin{equation} \label{eq:MSE2D}
\begin{split}
\E[X^2] = & \p^2 + \frac{\p(1-\p)}{\N^2} \times \\%
& \Bigg(\N +4 \sum_{i=1}^{\sqrt{\N}-1}\sum_{j=0}^{\sqrt{\N}-1}(\sqrt{\N}-i)(\sqrt{\N}-j) \\%
& \Max(0, (1-i\nd))\Max(0, (1-j\nd))\Bigg).
\end{split}
\end{equation}

Thus, $V=1-\E[X^2]$ leads to \ref{eq:visbility}.

\section*{Acknowledgment}
%
The authors would like to thank Bettina Gruen of Johannes Kepler University Linz for her help on deriving \ref{eq:visbility}.
This research was funded by the Austrian Science Fund (FWF) under grant number P 32185-NBL.
\ifCLASSOPTIONcaptionsoff
  \newpage
\fi
\bibliographystyle{IEEEtran}
\bibliography{references}

\begin{thebibliography}{10}
\providecommand{\url}[1]{#1}
\csname url@samestyle\endcsname
\providecommand{\newblock}{\relax}
\providecommand{\bibinfo}[2]{#2}
\providecommand{\BIBentrySTDinterwordspacing}{\spaceskip=0pt\relax}
\providecommand{\BIBentryALTinterwordstretchfactor}{4}
\providecommand{\BIBentryALTinterwordspacing}{\spaceskip=\fontdimen2\font plus
\BIBentryALTinterwordstretchfactor\fontdimen3\font minus
  \fontdimen4\font\relax}
\providecommand{\BIBforeignlanguage}[2]{{%
\expandafter\ifx\csname l@#1\endcsname\relax
\typeout{** WARNING: IEEEtran.bst: No hyphenation pattern has been}%
\typeout{** loaded for the language `#1'. Using the pattern for}%
\typeout{** the default language instead.}%
\else
\language=\csname l@#1\endcsname
\fi
#2}}
\providecommand{\BIBdecl}{\relax}
\BIBdecl

\bibitem{Moreira2013}
A.~Moreira, P.~Prats-Iraola, M.~Younis, G.~Krieger, I.~Hajnsek, and K.~P.
  Papathanassiou, ``A tutorial on synthetic aperture radar,'' \emph{IEEE
  Geoscience and Remote Sensing Magazine}, vol.~1, no.~1, pp. 6--43, March
  2013.

\bibitem{Li2015}
C.~J. Li and H.~Ling, ``Synthetic aperture radar imaging using a small consumer
  drone,'' in \emph{2015 IEEE International Symposium on Antennas and
  Propagation USNC/URSI National Radio Science Meeting}, July 2015, pp.
  685--686.

\bibitem{Rosen2000}
P.~A. Rosen, S.~Hensley, I.~R. Joughin, F.~K. Li, S.~N. Madsen, E.~Rodriguez,
  and R.~M. Goldstein, ``Synthetic aperture radar interferometry,''
  \emph{Proceedings of the IEEE}, vol.~88, no.~3, pp. 333--382, March 2000.

\bibitem{Levanda2010}
R.~Levanda and A.~Leshem, ``Synthetic aperture radio telescopes,'' \emph{Signal
  Processing Magazine, IEEE}, vol.~27, pp. 14 -- 29, 02 2010.

\bibitem{Dravins2015}
D.~Dravins, T.~Lagadec, and P.~D. Nuñez, ``Optical aperture synthesis with
  electronically connected telescopes.'' \emph{Nature communications}, vol.~6,
  p. 6852, Apr 2015.

\bibitem{Ralston2007}
T.~S. Ralston, D.~L. Marks, P.~S. Carney, and S.~A. Boppart, ``{Interferometric
  synthetic aperture microscopy (ISAM)},'' \emph{Nature Physics}, pp.
  965--1004, 2007.

\bibitem{Hayes2009}
M.~P. Hayes and P.~T. Gough, ``Synthetic aperture sonar: a review of current
  status,'' \emph{IEEE Journal of Oceanic Engineering}, vol.~34, no.~3, pp.
  207--224, 2009.

\bibitem{Hansen2011}
\BIBentryALTinterwordspacing
R.~E. Hansen, ``Introduction to synthetic aperture sonar,'' in \emph{Sonar
  Systems Edited}.\hskip 1em plus 0.5em minus 0.4em\relax InTech Published,
  2011. [Online]. Available: \url{http://www.intechopen.com/books/sonar-
  systems/introduction-to-synthetic-aperture-sonar}
\BIBentrySTDinterwordspacing

\bibitem{Jensen2006Review}
\BIBentryALTinterwordspacing
J.~A. Jensen, S.~I. Nikolov, K.~L. Gammelmark, and M.~H. Pedersen, ``Synthetic
  aperture ultrasound imaging,'' \emph{Ultrasonics}, vol.~44, pp. e5 -- e15,
  2006, proceedings of Ultrasonics International (UI’05) and World Congress
  on Ultrasonics (WCU). [Online]. Available:
  \url{http://www.sciencedirect.com/science/article/pii/S0041624X06003374}
\BIBentrySTDinterwordspacing

\bibitem{Zhang2016}
\BIBentryALTinterwordspacing
H.~K. Zhang, A.~Cheng, N.~Bottenus, X.~Guo, G.~E. Trahey, and E.~M. Boctor,
  ``Synthetic tracked aperture ultrasound imaging: design, simulation, and
  experimental evaluation,'' \emph{Journal of medical imaging (Bellingham,
  Wash.)}, vol.~3, no. 27088108, pp. 027\,001--027\,001, Apr. 2016. [Online].
  Available: \url{https://www.ncbi.nlm.nih.gov/pmc/PMC4824841/}
\BIBentrySTDinterwordspacing

\bibitem{Barber2014}
Z.~W. Barber and J.~R. Dahl, ``Synthetic aperture ladar imaging demonstrations
  and information at very low return levels.'' \emph{Applied optics}, vol.~53,
  pp. 5531--5537, Aug 2014.

\bibitem{Turbide2017}
\BIBentryALTinterwordspacing
S.~Turbide, L.~Marchese, M.~Terroux, and A.~Bergeron, ``Synthetic aperture
  lidar as a future tool for earth observation,'' \emph{Proc.SPIE}, vol. 10563,
  pp. 10\,563 -- 10\,563 -- 8, 2017. [Online]. Available:
  \url{https://doi.org/10.1117/12.2304256}
\BIBentrySTDinterwordspacing

\bibitem{Vaish04}
V.~Vaish, B.~Wilburn, N.~Joshi, and M.~Levoy, ``Using plane + parallax for
  calibrating dense camera arrays,'' in \emph{Proc. CVPR}, 2004, pp. 2--9.

\bibitem{Vaish06}
V.~{Vaish}, M.~{Levoy}, R.~{Szeliski}, and C.~L.~Z. and, ``Reconstructing
  occluded surfaces using synthetic apertures: Stereo, focus and robust
  measures,'' in \emph{2006 IEEE Computer Society Conference on Computer Vision
  and Pattern Recognition (CVPR'06)}, vol.~2, June 2006, pp. 2331--2338.

\bibitem{Zhang2018}
H.~Zhang, X.~Jin, and Q.~Dai, ``Synthetic aperture based on plenoptic camera
  for seeing through occlusions,'' in \emph{Advances in Multimedia Information
  Processing -- PCM 2018}, R.~Hong, W.-H. Cheng, T.~Yamasaki, M.~Wang, and
  C.-W. Ngo, Eds.\hskip 1em plus 0.5em minus 0.4em\relax Cham: Springer
  International Publishing, 2018, pp. 158--167.

\bibitem{YangTao2016}
\BIBentryALTinterwordspacing
T.~Yang, W.~Ma, S.~Wang, J.~Li, J.~Yu, and Y.~Zhang, ``Kinect based real-time
  synthetic aperture imaging through occlusion,'' \emph{Multimedia Tools and
  Applications}, vol.~75, no.~12, pp. 6925--6943, Jun 2016. [Online].
  Available: \url{https://doi.org/10.1007/s11042-015-2618-1}
\BIBentrySTDinterwordspacing

\bibitem{Joshi2007}
N.~{Joshi}, S.~{Avidan}, W.~{Matusik}, and D.~J. {Kriegman}, ``Synthetic
  aperture tracking: Tracking through occlusions,'' in \emph{2007 IEEE 11th
  International Conference on Computer Vision}, Oct 2007, pp. 1--8.

\bibitem{ZhaoPei2019}
Z.~Pei, Y.~Li, M.~Ma, J.~Li, C.~Leng, X.~Zhang, and Y.~Zhang, ``Occluded-object
  3d reconstruction using camera array synthetic aperture imaging,''
  \emph{Sensors}, vol.~19, p. 607, 01 2019.

\bibitem{YangTao2014}
T.~Yang, Y.~Zhang, J.~Yu, J.~Li, W.~Ma, X.~Tong, R.~Yu, and L.~Ran,
  ``All-in-focus synthetic aperture imaging,'' in \emph{Computer Vision -- ECCV
  2014}, D.~Fleet, T.~Pajdla, B.~Schiele, and T.~Tuytelaars, Eds.\hskip 1em
  plus 0.5em minus 0.4em\relax Cham: Springer International Publishing, 2014,
  pp. 1--15.

\bibitem{ZhaoPei2013}
\BIBentryALTinterwordspacing
Z.~Pei, Y.~Zhang, X.~Chen, and Y.-H. Yang, ``Synthetic aperture imaging using
  pixel labeling via energy minimization,'' \emph{Pattern Recognition},
  vol.~46, no.~1, pp. 174 -- 187, 2013. [Online]. Available:
  \url{http://www.sciencedirect.com/science/article/pii/S0031320312002841}
\BIBentrySTDinterwordspacing

\bibitem{Kurmi2018}
I.~Kurmi, D.~Schedl, and O.~Bimber, ``Airborne optical sectioning,'' \emph{J.
  Imaging}, vol.~4, no. 102, 2018.

\bibitem{Bimberi2019IEEECGA}
O.~Bimber, I.~Kurmi, and D.~Schedl, ``Synthetic aperture imaging with drones,''
  \emph{IEEE Computer Graphics and Applications}, 2019.

\bibitem{Levoy1996}
M.~Levoy and P.~Hanrahan, ``Light {F}ield {R}endering,'' in \emph{Proceedings
  of the 23rd Annual Conference on Computer Graphics and Interactive
  Techniques}, ser. SIGGRAPH '96.\hskip 1em plus 0.5em minus 0.4em\relax New
  York, NY, USA: ACM, 1996, pp. 31--42.

\bibitem{Isaksen2000}
A.~Isaksen, L.~McMillan, and S.~J. Gortler, ``Dynamically reparameterized light
  fields,'' in \emph{Proceedings of the 27th annual conference on Computer
  graphics and interactive techniques}, ser. SIGGRAPH '00.\hskip 1em plus 0.5em
  minus 0.4em\relax ACM Press/Addison-Wesley Publishing Co.,, 2000, pp.
  297--306.

\bibitem{SignalAveraging2010}
\BIBentryALTinterwordspacing
U.~Hassan and M.~S. Anwar, ``Reducing noise by repetition: introduction to
  signal averaging,'' \emph{European Journal of Physics}, vol.~31, no.~3, pp.
  453--465, 2010. [Online]. Available:
  \url{https://doi.org/10.1088%2F0143-0807%2F31%2F3%2F003}
\BIBentrySTDinterwordspacing

\bibitem{Trochta2017}
\BIBentryALTinterwordspacing
J.~Trochta, M.~Krůček, T.~Vrška, and K.~Král, ``3d forest: An application
  for descriptions of three-dimensional forest structures using terrestrial
  lidar,'' \emph{PLOS ONE}, vol.~12, no.~5, pp. 1--17, 05 2017. [Online].
  Available: \url{https://doi.org/10.1371/journal.pone.0176871}
\BIBentrySTDinterwordspacing

\bibitem{wang2004image}
Z.~Wang, A.~C. Bovik, H.~R. Sheikh, and E.~P. Simoncelli, ``Image quality
  assessment: from error visibility to structural similarity,'' \emph{{IEEE}
  Trans. Image Process.}, vol.~13, no.~4, pp. 600--612, 2004.

\end{thebibliography}
%
%
\end{document}